\documentclass[prl,preprint,superscriptaddress]{revtex4-1}
\usepackage{amsmath}
\usepackage{amsfonts}
\usepackage{amssymb}
\usepackage{graphicx}

\newcommand{\dif}{\mathrm{d}}

\renewcommand \theequation {S\arabic{equation}}%
\renewcommand \thefigure {S\arabic{figure}}

\begin{document}
\title{Graphene on silicon nitride for optoelectromechanical micromembrane resonators}

\author{Silvan Schmid}
\affiliation{Department of Micro- and Nanotechnology, Technical University of Denmark, DTU Nanotech, DK-2800 Kongens Lyngby, Denmark}
\author{Tolga Bagci}
\author{Emil Zeuthen}
\affiliation{QUANTOP, Niels Bohr Institute, University of Copenhagen, DK-2100 Copenhagen, Denmark}
\author{Jacob M. Taylor}
\affiliation{Joint Quantum Institute/NIST, College Park, Maryland, USA}
\author{Patrick K. Herring}
\author{Maja C. Cassidy}
\affiliation{School of Engineering and Applied Science, Harvard University, Cambridge, Massachusetts 02138, USA}
\author{Charles M. Marcus}
\affiliation{Center for Quantum Devices, Niels Bohr Institute, University of Copenhagen, DK-2100, Denmark}
\author{Luis Guillermo Villanueva}
\author{Bartolo Amato}
\author{Anja Boisen}
\affiliation{Department of Micro- and Nanotechnology, Technical University of Denmark, DTU Nanotech, Building 345 East, DK-2800 Kongens Lyngby, Denmark}
\author{Yong Cheol Shin}
\author{Jing Kong}
\affiliation{Department of Materials Science and Engineering, Massachusetts Institute of Technology, Cambridge, Massachusetts 02139, USA}
\author{Anders S. S{\o}rensen}
\author{Koji Usami}
\affiliation{QUANTOP, Niels Bohr Institute, University of Copenhagen, DK-2100 Copenhagen, Denmark}
\author{Eugene S. Polzik}
\affiliation{QUANTOP, Niels Bohr Institute, University of Copenhagen, DK-2100 Copenhagen, Denmark}

\begin{abstract}
Due to their exceptional mechanical and optical properties, dielectric silicon nitride (SiN) micromembrane resonators have become the centerpiece of many optomechanical experiments. Efficient capacitive coupling of the membrane to an electrical system would facilitate exciting hybrid optoelectromechanical devices. However, capacitive coupling of such dielectric membranes is rather weak. Here we add a single layer of graphene on SiN micromembranes and compare electromechanical coupling and mechanical properties to bare dielectric membranes and to membranes metallized with an aluminium layer. The electrostatic coupling of graphene coated membranes is found to be equal to a perfectly conductive membrane. Our results show that a single layer of graphene substantially enhances the electromechanical capacitive coupling without significantly adding mass, decreasing the superior mechanical quality factor or affecting the optical properties of SiN micromembrane resonators.
\end{abstract}

\maketitle
%\section{Introduction}
Hybrid devices capable of coupling different systems are presently one of the hot topics in quantum technologies \cite{Teufel2011, Gavartin2012}. Recent proposals \cite{Taylor2011,Regal2011} outline optoelectromechanical systems, where a mechanical resonator is strongly coupled to an optical and an electrical resonator at the same time. The centerpiece of these proposals is a mechanical micromembrane inside an optical resonator which is capacitively coupled  to an LC circuit \cite{Taylor2011}. %A key figure of merit is the strength of the electromechanical coupling. In this work, we deposit a single layer of graphene onto SiN membranes (SiN-G) and compare their mechanical properties and electromechanical coupling to bare SiN membranes and to membranes covered with an aluminium layer (SiN-Al) \cite{Yu2012}.
Recently, such a hybrid optoelectromechanical system has been realized based on SiN-Al membranes \cite{Bagci2013}. For these applications it is essential to have a strong electromechanical coupling relative to the mechanical properties of the membrane.

A key feature of SiN micromembranes is their ultrahigh quality factor (Q) reaching $10^6 - 10^7$, low mass and excellent optical transparency with losses less than $10^{-5}$ in the near infrared \cite{thompson2008strong, wilson2009cavity, Kippenberg2007}. In order to not downgrade these essential properties it was proposed to use dielectric polarization forces on bare SiN membranes for the electromechanical coupling \cite{Taylor2011}. Such forces on a dielectric are however, inherently weaker than the forces on conductors. In this work, we deposit a single layer of graphene onto SiN membranes (SiN-G) \cite{Lee2013} and compare their mechanical properties and electromechanical coupling to bare SiN membranes and to membranes covered with an aluminium layer (SiN-Al) \cite{Yu2012}.
We show that this single layer of graphene allows superior electromechanical coupling without deteriorating the exceptional properties of SiN membranes. Importantly, our setup uses graphene in a floating electrode configuration. Thereby graphene is not in contact with a metal and the interaction happens electrostatically. In this fashion we avoid the large contact resistance associated with connecting graphene and a metal electrode \citep{Russo2010,Venugopal2010,Novoselov2012}.

\begin{figure}
\centering
\includegraphics[width=0.5\textwidth]{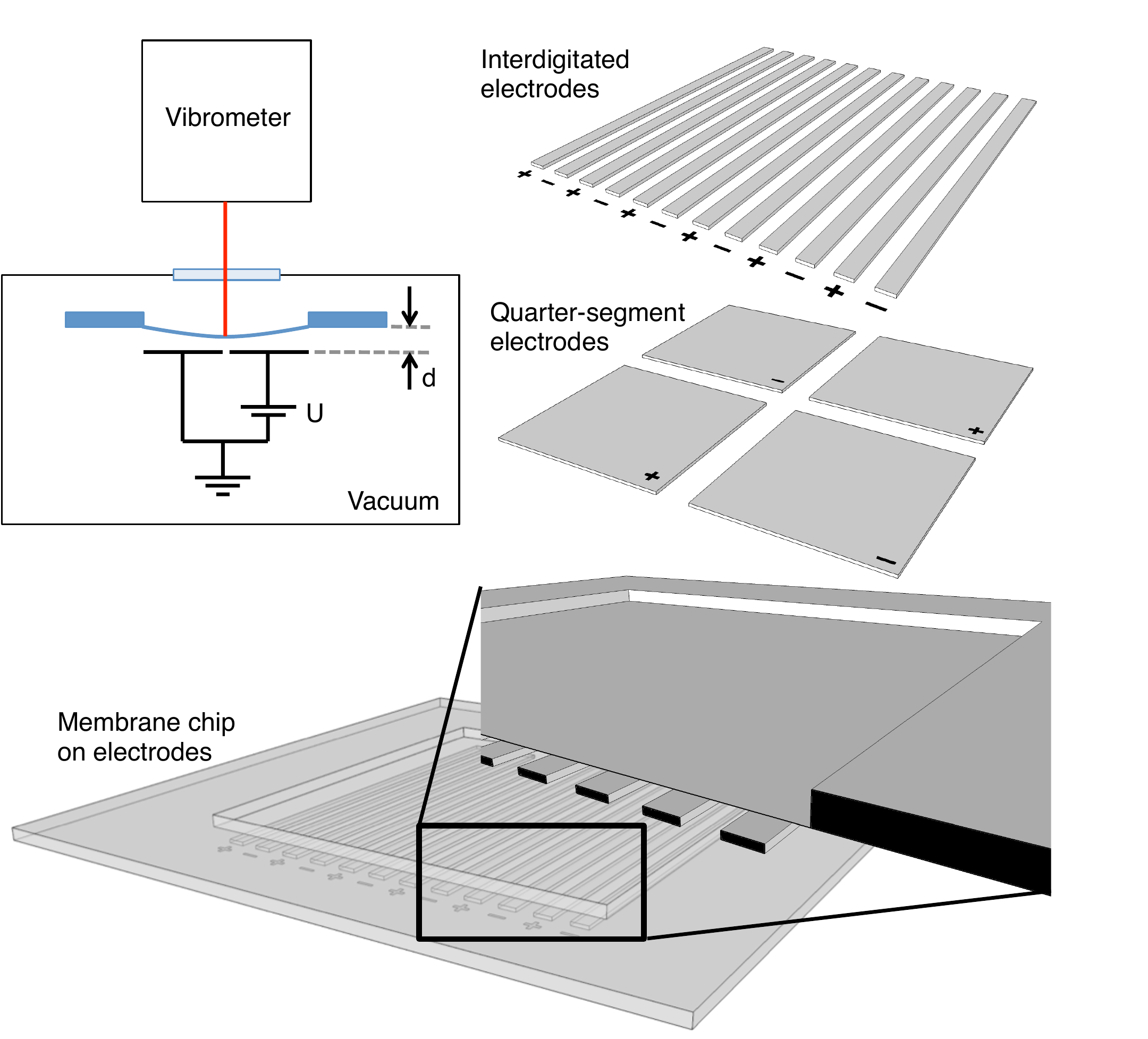}
\caption{Schematic drawing of the experimental setup. The micromechanical membranes are placed on top of coplanar electrodes. Two types of electrodes are used: interdigited finger electrodes and quarter-segment electrodes.}
\label{fig:fig1}
\end{figure}

We used commercial high and low-stress 50~nm thick Si$_3$N$_4$ membranes (Norcada Inc.) for both the SiN and SiN-G resonators. Single layer graphene was grown on copper foil using standard CVD techniques \cite{Li2009}. The graphene on copper was cut to size, the copper wet-etched and the graphene transferred to the surface of the membrane in the aqueous environment. A thin layer of PMMA was used to support the graphene during the etching and transfer process. Once dry, the PMMA layer was removed from the graphene using acetone vapor.  The final structure was robust enough for subsequent fabrication steps, for instance for making a round opening window in the graphene for future optical cavity applications or lithographically defining gates for control of the carrier density and type. In the  experiments described below we used membranes without the opening in the graphene layer. A characterization by Raman spectroscopy of the single layer graphene on SiN-G chips is presented in the supplementary information. The Al covered membranes were fabricated in-house by standard cleanroom processing. The high-stress stoichiometric SiN layer is 100~nm thick. The aluminium layer is 50~nm thick and it is patterned by a lift-off step. The Al layer could also be fabricated with a round hole in the center of the membrane for optical access. A small rim (5\% of the membrane size) along the anchor of the membrane was spared out in order to minimize damping \citep{Yu2012,schmid2011damping}.

Two types of coplanar electrode chips were used: interdigitated electrodes and quarter-segment electrodes. Schematic drawings of two types of electrodes are depicted in Fig.~\ref{fig:fig1}. The electrodes fabricated by standard cleanroom processing are made of a 200~nm thick gold layer sitting on borosilicate glass substrate or SiN covered silicon substrate. The membranes are placed membrane downwards onto the electrode. The electrode chips feature pillars with a height of 600~nm and 1~$\mu$ in order to define a small gap between membrane and electrodes. Optical measurements of the gap distance $d$ and the membrane vibrations have been made with a white light interferometer (vibrometer MSA-500 from Polytec GmbH). The gap distance ranged from 3.5 to 14~$\mu$m which is larger than the height of the dedicated pillars. The larger measured distance can be ascribed to electrode chip unevenness coming from dirt and chip stress gradients and has been reduced down to the pillar height size in subsequent experiments.

The experimental setup is schematically depicted in Fig.~\ref{fig:fig1}. The membrane-electrode sandwich is placed in a vacuum chamber (pressure below $1\times10^{-5}$~mbar) and is electrically connected to an external voltage source.  For quality factor measurements, the membranes were placed on a piezo for stimulation. The quality factors were extracted from the membrane ring-down time and the -3dB bandwidth of the resonance peak.

\begin{figure}
\centering
\includegraphics[width=0.45\textwidth]{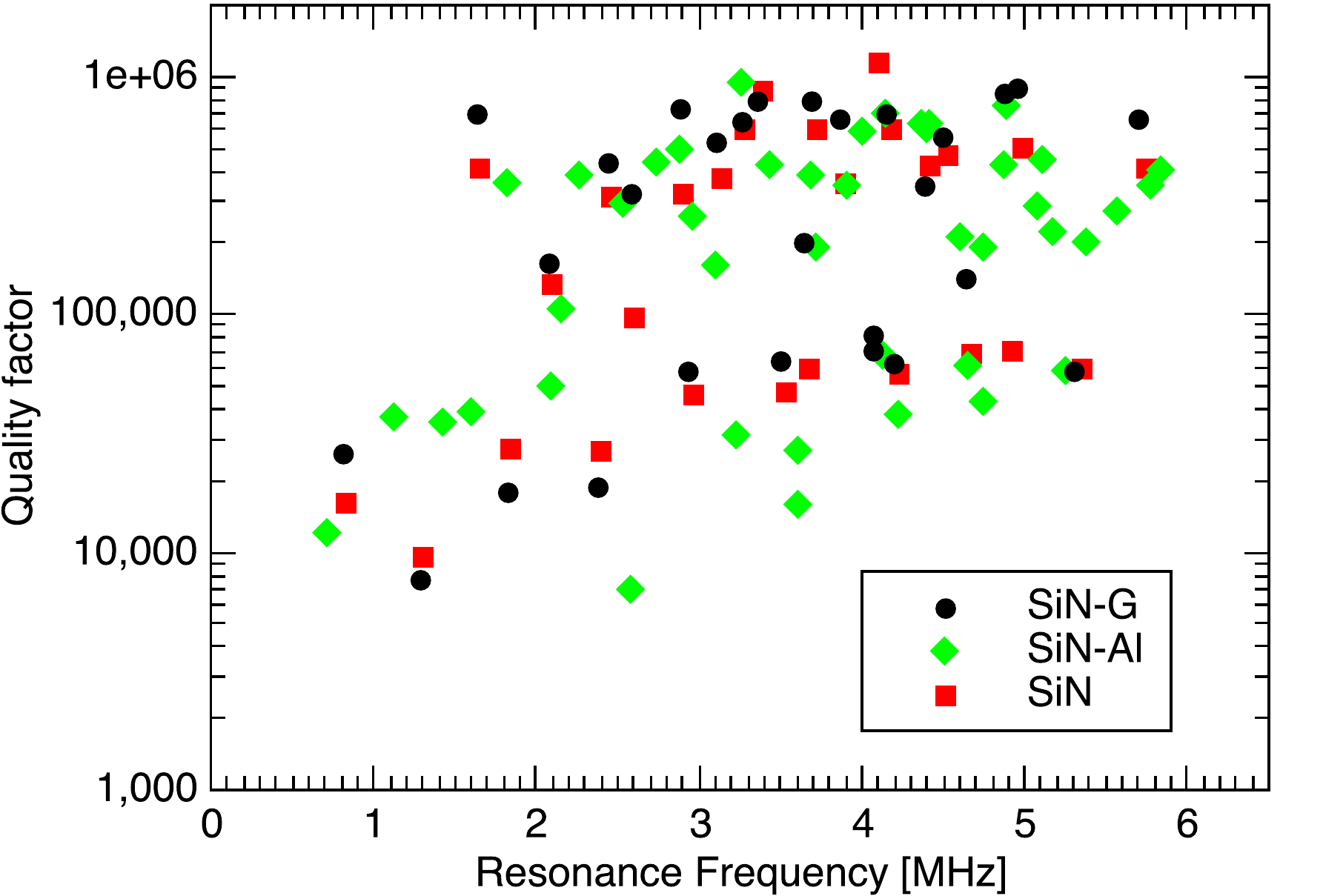}
\caption{Quality factors with increasing mode numbers of a SiN-G membrane compared to a bare SiN and a SiN-Al membrane. The membranes have a diameter of 0.5$\times$0.5~mm$^2$ and are all made of high-stress stoichiometric SiN.}
\label{fig:fig2}
\end{figure}

%Mechanical behavior
A key feature of micromechanical membranes is their Q of up to several million. Such high quality factors are a requirement for strong optomechanical coupling and for  high resolution measurements. %It has been shown that a metal coating increases the energy loss in a membrane which results in a reduction of Q. The contribution of the mechanical dissipation of the metal layer can be minimized by leaving the membrane area close to the clamping, where the maximal mechanical strain occurs, uncoated \citep{Yu2012}. 
%The hypothesis is that a single graphene layer will not significantly contribute to the energy dissipation of a SiN membrane. 
In the first experiment we compared the quality factors of the 3 different types of membranes. Fig.~\ref{fig:fig2} shows the measured quality factors of a bare SiN, SiN-Al, and SiN-G membrane. On the SiN-Al membrane a rim of 5\% of the membrane dimension remained uncoated, whereas graphene is fully covering the entire SiN-G membrane. The quality factors are highly mode dependent and the measured values are in correspondence with values measured by Yu et al. \cite{Yu2012}. There are clearly two sets of Q-values. According to \cite{Yu2012}, the lower set (below $\sim100000$) is limited by clamping losses. The higher set is limited by intrinsic damping, such as bulk or surface losses. It can be seen from the measurements that there is no significant difference between the different membrane types in both sets. The graphene sheet seems to be mechanically invisible and it does not significantly contribute to the energy loss. Thus with graphene it is not required to spare out the membrane area close to the clamping edge as it is for metal layers. The additional metal layer of the SiN-Al membrane downshifts the resonance frequencies compared to the other two membranes, as can be clearly seen in Fig.~\ref{fig:fig2} for low mode numbers. This divergence increases for higher mode numbers, which makes a direct comparison difficult.
%It is nevertheless noteworthy that the Q values measured with the SiN-Al membrane are comparable to the other two membranes. It proofs that high Q values can be achieved with metal coated %membranes by leaving the edges blank.\cite{Yu2012}.

It is crucial for an efficient hybrid optoelectromechanical device to have strong electromechanical coupling between the membrane and the electrical circuit. The electrostatic interaction for SiN membranes is due to dielectric polarization forces (the electric field interacts with a dielectric) \citep{Taylor2011, Schmid2006, Schmid2010, Unterreithmeier2009} and in some of the experiments presented here it is also due to the quasi-permanent electric charges in a dielectric. The interdigitated electrodes have been design to generate strong electric field gradients that are required for the dielectric polarization force. The interaction for conductive SiN-Al \citep{Teufel2011} is due to the electrostatic force between conductors (the electric field interacts with charges on a conductor). For this case the quarter-segment electrodes have been designed. Using the two electrode geometries allows us to characterize the comparative performance of both types of membranes. It also allows us to conclude to which type the SiN-G membranes belong.

Neglecting the effect of free charges on the membrane, the electrostatic force between a dielectric or conductive thin membrane and electrodes can generally be described by
\begin{equation}\label{eq:force}
F=c A f(d) U^2
\end{equation}
with $A$ - the membrane area, $U$ - the potential difference over the electrodes, $c$ - the electrostatic force constant that characterizes the coupling performance, and $f(d)$ - a function describing the distance dependence of the force.

We  extract the intrinsic coupling strength $c$ of each membrane type from measuring the mechanical frequency shift due to the so-called spring softening effect which results in a quadratic frequency drop with the DC-bias voltage $U_{DC}$. The relative frequency shift is given by (see supplementary information)
\begin{equation}\label{eq:springsoftening}
\frac{\Delta\omega}{\omega_0}= - \alpha U_{DC}^2.
\end{equation}
Using $\alpha$ found from the frequency shift data (\ref{eq:springsoftening}), the electrostatic force constant $c$ can be calculated from
\begin{equation}\label{eq:forceconstant}
c = 2 \alpha [-f'(d)]^{-1} h \rho\omega_0^2 \eta^{-1}
\end{equation} 
with the membrane thickness $h$ and the membrane mass density $\rho$; while $\eta$ (of order unity) quantifies the spatial overlap between the membrane mode shape and the fixed electrodes, in particular compensating for the gap between quarter-segment electrodes and for the circular hole in the SiN-Al membranes' center (see supplementary information).

Here the coupling performance of SiN-G membranes is compared to that of the SiN and SiN-Al membranes using the two types of capacitor+membrane chips described above. In the first set, SiN and SiN-G membranes are coupled to interdigitated electrodes; in the second set, SiN-Al and SiN-G membranes are coupled to quarter-segment electrodes.

%First we consider the scenario of using dielectric polarization forces for the electromechanical coupling of a bare %nonconductive SiN membrane \citep{Taylor2011, Schmid2010}. This force is a function of the electric field gradient. In %order to generate a strong electric field gradient, the coplanar electrodes with thin interdigitated fingers are used.

For the interdigitated electrode geometry and the range of membrane-capacitor distances $d$ used in the experiments, the electrostatic force for both dielectric and perfectly conducting membranes (an appropriate model for our SiN-G membranes) is well approximated by
\begin{equation}\label{eq:forcelaw-inter}
f(d)=A_{0}^{-1}e^{-\kappa d}
\end{equation}
with $A_{0}$ - a scaling constant with units of area. For the data presented below, we use $A_0 = 1$~$\mu$m$^2$ For our specific setup, we determined numerically that $\kappa=1.05\mu m^{-1}$ for both dielectric and perfectly conducting membranes.

 %The Eq.~(\ref{eq:force_IDC}) can be cast in the generic form.
%Hence, for the interdigitated electrodes we use the following model for the electrostatic force on a patch of membrane of %area $A$ (in absence of quasi-static charges):
%where $U$ is the voltage difference between the positive and negative fixed electrodes, and $c$ is an electrostatic force %constant that characterizes the coupling performance of the material ($A_{0}$ is an arbitrary scaling constant with units %of area).

\begin{figure}
\centering
\includegraphics[width=0.45\textwidth]{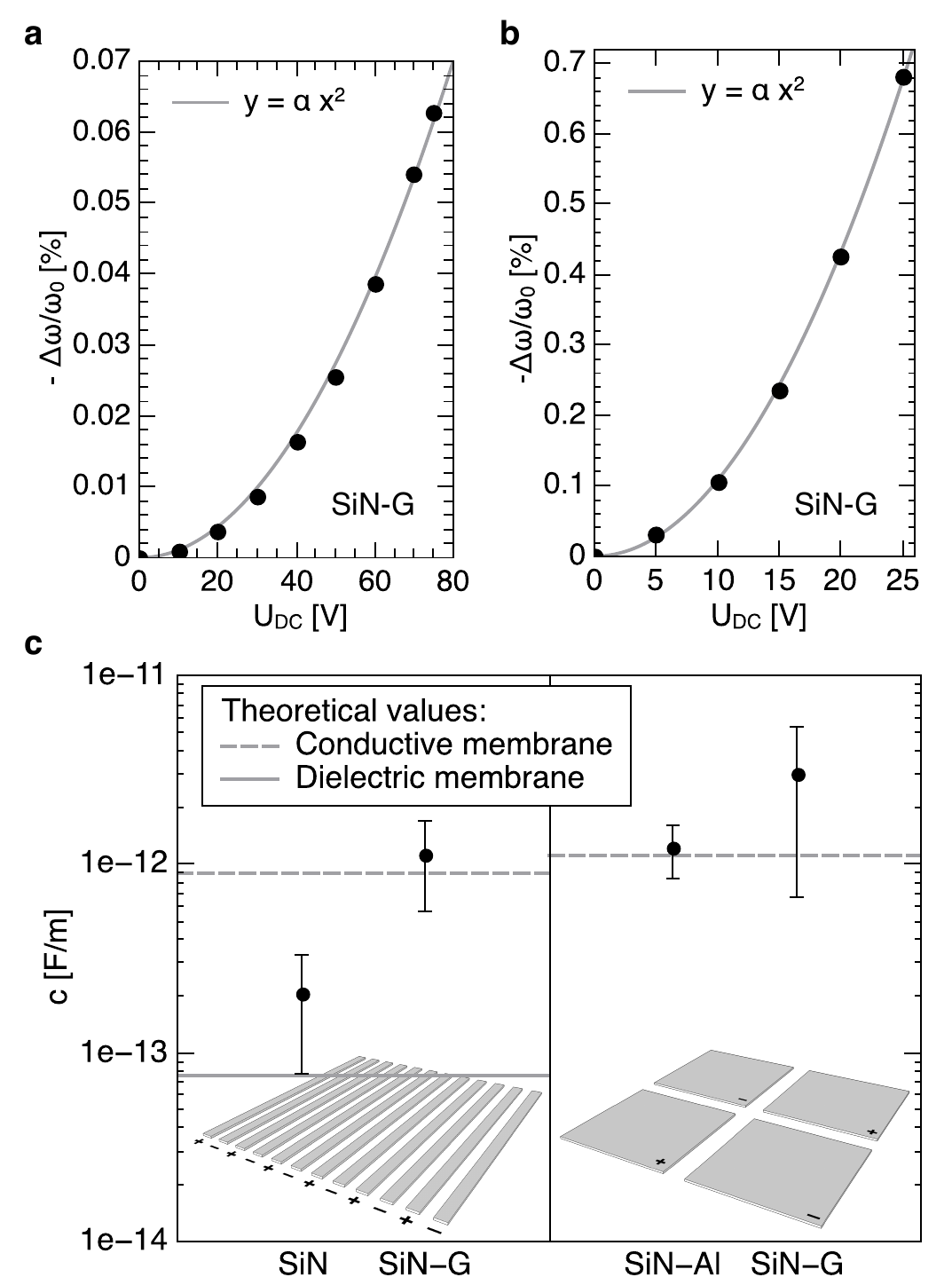}
\caption{Extraction and comparison of the electrostatic force constant $c$. a) Resonance frequency of fundamental mode of a $0.5\times0.5$~mm$^2$ SiN-G membrane on interdigitated electrodes as a function of DC voltage. The distance between membrane and electrodes is $d=7.0$~$\mu$m. b) Resonance frequency fundamental mode of a $0.5\times0.5$~mm$^2$ SiN-G membrane on quarter-segment electrodes as a function of DC voltage. The distance between the membrane and electrodes is $d=6.3$~$\mu$m. c) Comparison of force constants $c$ (according to (\ref{eq:forceconstant}) using (\ref{eq:forcelaw-inter}) and (\ref{eq:forcelaw-seg})) for different combinations of electrodes and membranes. On the left, $c$ extracted for bare SiN (4 experiments) and SiN-G (3 experiments) membranes on interdigitated electrodes are shown. The electrode fingers are 4~$\mu$m wide with a gap of 2~$\mu$m between the fingers. To the right, $c$ of SiN-Al (3 experiments) and SiN-G (4 experiments) on quarter-segment electrodes are shown, with a gap between the segments of 60~$\mu$m. The error bars represent the standard deviation of the conducted experiments. The solid and dashed grey lines represents the theoretical values for the scenario of a pure dielectric polarization force and an electrostatic force on the conductive membrane, respectively.}
\label{fig:c}
\end{figure}

In Fig.~\ref{fig:c}a the electrostatic spring softening of a SiN-G membrane on top of an interdigitated electrode is shown. From the fit parameter $\alpha$ obtained from this data the force constant $c$ is calculated for the membranes using Eq. (\ref{eq:forceconstant}) and the exponential force law (\ref{eq:forcelaw-inter}). The resulting average force constants for the bare SiN and SiN-G membranes on interdigitated electrodes are shown in the left part of Fig.~\ref{fig:c}c. SiN-G is seen to outperform SiN by a factor of $~5.5$. Also shown are the theoretical values %based on an analytical model \cite{Schmid2010} (see supplementary information).
determined semi-analytically for dielectric (SiN, $\varepsilon_{\text{r}}=7.6$) and perfectly conducting membranes; the average force constant of SiN-G is seen to be compatible with the latter.

% Quarter-segment electrodes
%********************************************

Next, we use the quarter-segment coplanar electrodes with a SiN-Al or a SiN-G membrane as a floating electrode. The results of this set of experiments are well described by another force law (in the regime where $d$ is smaller than the inter-electrode gap):
\begin{equation}\label{eq:forcelaw-seg}
f(d)=1/d^2.
\end{equation}

The measured spring softening of a SiN-G membrane placed over quarter segment electrodes is shown in Fig.~\ref{fig:c}b. From the fit parameter $\alpha$ obtained from this data the force constant $c$  is calculated for the conducting membranes again with Eq.~(\ref{eq:forceconstant}) but now using the quadratic force law (\ref{eq:forcelaw-seg}). On the right side of Fig.~\ref{fig:c}c the extracted average force constants of SiN-G and SiN-Al membranes are shown. The experimental values again agree well with the theoretical value for a perfect conductor. The large standard deviation of the force constant can be assigned to uncertainties in the distance $d$ measurement and lateral misalignment which can contribute up to 20\% error. The relatively large average $c$ value of the SiN-G membranes might be an effect due to the excess graphene on the frame, as discussed in the supplementary information.

%It can be seen that the relative frequency shift for a similar distance $d$ of SiN-G membranes on  quarter-segment electrodes (Fig.~\ref{fig:c}b) is approximately two orders of magnitude higher than on interdigitated electrodes (Fig.~\ref{fig:c}a). The absolute electromechanical coupling strength of quarter-segment electrodes is less distance dependent (quadratic vs. exponential force law) and therefore outperforms the coupling of interdigitated electrodes (for the separations $d$ considered here).

\begin{figure}
\centering
\includegraphics[width=0.5\textwidth]{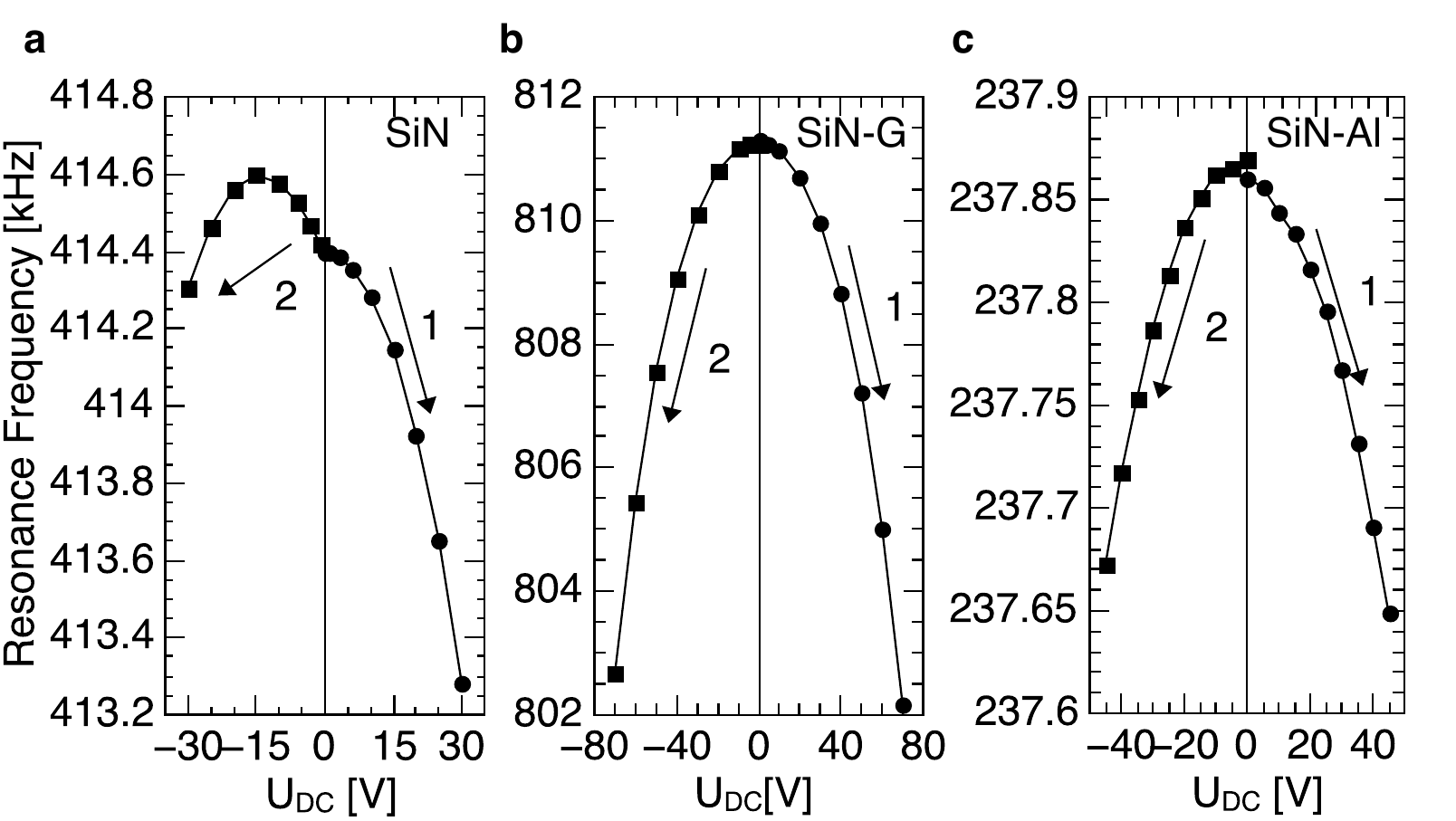}
\caption{Response of the resonance frequency of fundamental mode of a bare SiN, a SiN-G, and a SiN-Al membrane to DC polarity inversion. First the voltage was increased in the positive direction. Then the polarity was reversed and the voltage was increased in the negative direction. a) Response of a bare SiN membrane ($1\times1$~mm$^2$, $d=4.0\mu$m) on interdigitated electrodes; electrode fingers are 4~$\mu$m wide with a gap of 2~$\mu$m between the fingers. b) Response of a SiN-G membrane ($0.5\times0.5$~mm$^2$, $d=5.5\mu$m) on interdigitated electrodes; the electrode fingers are 4~$\mu$m wide with a gap of 5~$\mu$m between the fingers. c) Response of a SiN-Al membrane ($1\times1$~mm$^2$, $d=11$~$\mu$m) on quarter-segment electrodes. The resonance frequency was determined from the thermomechanical resonance peak.}
\label{fig:charging}
\end{figure}

An additional complication with SiN membranes which is eliminated by using SiN-Al or SiN-G membranes is the accumulation of free charges on the dielectric SiN membrane. The charging has been investigated by increasing the DC voltage and then inverting the polarity. If quasi-static charges are present, a spring hardening (increase in oscillation frequency) instead of softening will be observed. The results of this experiment are shown in Fig.~\ref{fig:charging} with a bare SiN, a SiN-Al and a SiN-G membrane on interdigitated electrodes. From the observed spring hardening, shown in Fig.~\ref{fig:charging}a, it can be concluded that free electric charges are indeed available on the bare SiN membranes. We note that the force on a charge is a function of the inverse of the distance ($\propto 1/d$) and may therefore be partly responsible for the somewhat higher value of the coupling constant c for SiN shown in Fig.~\ref{fig:c}c. The charging of the bare SiN membrane is investigated in more detail in the supplementary information.

Fig.~\ref{fig:charging}b and  Fig.~\ref{fig:charging}c present the results of the same charging experiment conducted with a SiN-G on similar interdigitated electrodes and with SiN-Al membranes on quarter-segment electrodes, respectively. Assuming that the electrode design does not influence the charging effect, we conclude that the absence of spring hardening for SiN-G shows that the single layer of graphene as well as the Al layer eliminates charging.

\begin{figure}
\centering
\includegraphics[width=0.45\textwidth]{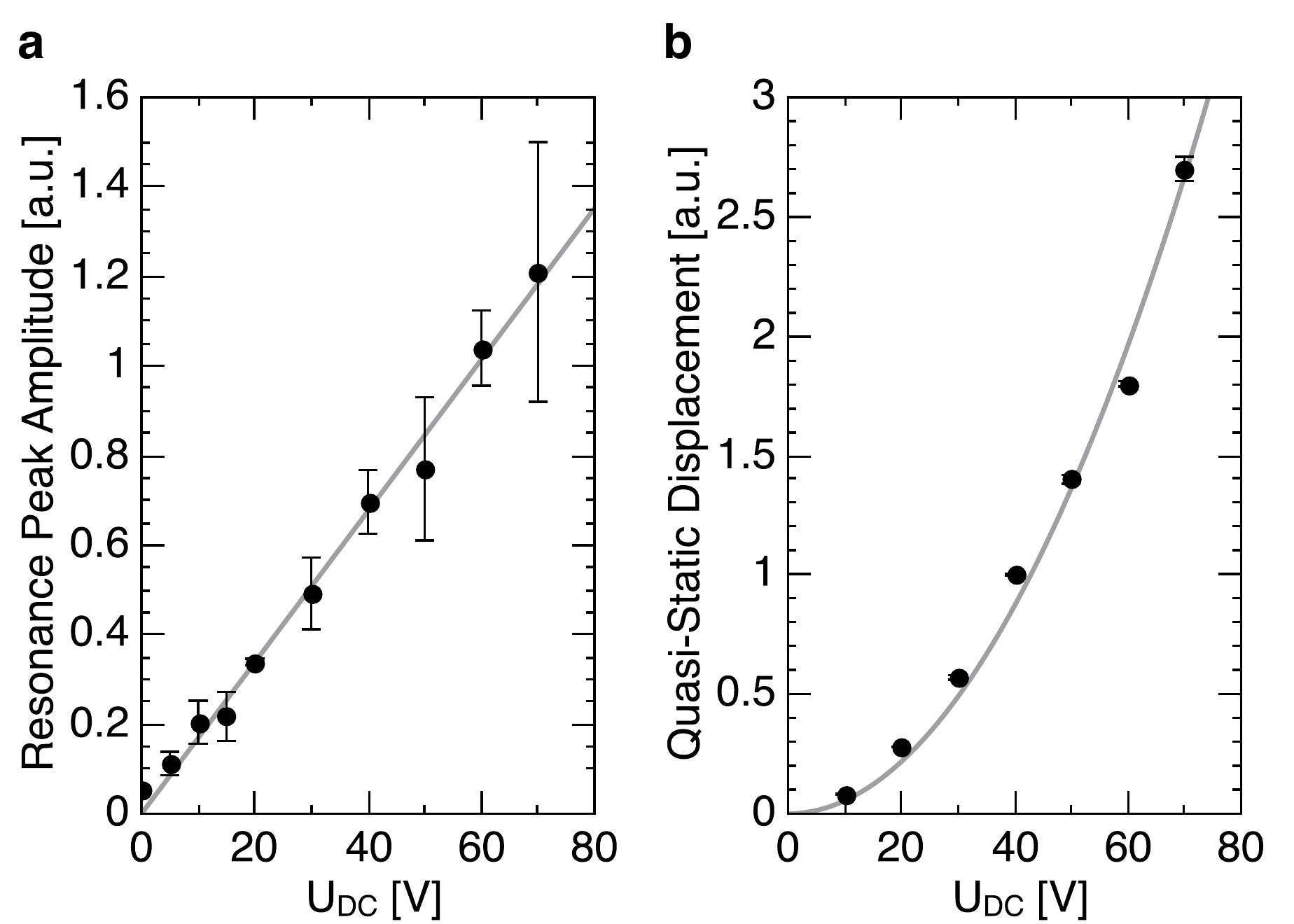}
\caption{Response of fundamental mode of a $0.5\times0.5$~mm$^2$ SiN-G membrane on interdigitated electrodes. Response of a) resonance peak amplitude and b) quasi-static displacement to DC bias voltage. The distance between membrane and electrodes is $d=6.5$~$\mu$m. The electrode fingers are 4~$\mu$m wide with a gap of 5~$\mu$m between the fingers. The resonance peak amplitude was measured for a white noise signal of $U_{AC}=10$~mV. The quasi-static displacement was measured with a rectangular signal at 10~kHz (which is far below the fundamental resonance frequency and can therefore be considered to be quasi-static). The error bars represent the standard deviation of 3 consecutive measurements. The grey lines represent a linear fit in a) and a quadratic fit in b).}
\label{fig:SiN-G}
\end{figure}

We performed further characterization of the SiN-G membranes by studying the dependence of the ac amplitude and the dc displacement on the coupling voltage (Fig.~\ref{fig:SiN-G}).
If the AC driving voltage is small compared to the DC bias voltage, the electrostatic force becomes
\begin{equation}
F = c A f(d) (U_{DC}+U_{AC})^2 \approx c A f(d) (U_{DC}^2 + 2 U_{AC}U_{DC}).
\end{equation}
Thus the mechanical resonance peak amplitude is approximately a linear function of the DC voltage, for a fixed AC driving voltage, and the static deflection of the membrane is a quadratic function of the DC voltage. As seen from Fig.~\ref{fig:SiN-G} the experimentally observed behavior supports the model.

%Fig.~\ref{fig:SiN-Al}b shows the mechanical resonance peak amplitude of a SiN-Al membrane as a function of negative and positive DC voltage. 

%We remark that in the case of additional capacitance between the electrodes and the excess graphene outside the membrane area, the theoretical force constant may be enhanced by up to a factor of 2 (if this capacitance is large and highly asymmetric between the two polarities). A similar enhancement may occur if charging of the membrane from one of the electrodes takes place on a timescale faster than the $U_{\text{DC}}$ bias scan time.

%Like the SiN-G membranes, the SiN-Al membranes did not show any signs of charging (see Fig.~\ref{fig:SiN-Al}). The SiN-G are behaving according to the theory, as can be seen from Fig.~\ref{fig:SiN-G}.

In conclusion we have shown that SiN-G membranes with a single layer of graphene are promising canditates for efficient optoelectromechanical hybrid coupling devices. Furthermore, by using the graphene sheet on SiN in a floating electrode configuration, we overcome the unsolved problem of high contact resistance between metal electrodes and graphene. They show electromechanical coupling which is as good as for an ideal conductor. The enhanced electromechanical coupling of conductive membranes lowers the threshold for strong coupling, as predicted for dielectric membranes \cite{Taylor2011}, accordingly. The single graphene layer is mechanically invisible and does not negatively influence the membrane performance.  Unlike the real metal coating, graphene does not require patterning in order to maintain the highest mechanical quality factors of a bare SiN membrane. Also, unlike the metal coating, graphene does not add any noticeable mass to the membrane which is an advantage for high sensitivity applications, such as, e.g., the optical detection of radio waves with metallized silicon nitride membrane resonators \cite{Bagci2013}. The lower mass achieved by SiN-G membranes as compared to SiN-Al membranes used in the reference would improve the quantum efficiency of RF-to-optical conversion by up to 4 times (by the ratio of the masses). Finally, the use of graphene overcomes the complications of charging effects in SiN membranes, which has shown to be difficult to control. 

%And finally, the absolute electromechanical coupling strength of quarter-segment electrodes is less distance dependent than interdigitated electrodes for both SiN-G and SiN-Al membranes. 

\vspace{0.5cm}
{\bf Acknowledgments}
\vspace{0.5cm}
We acknowledge funding from the DARPA program QUASAR, the EU project QESSENCE and the ERC projects INTERFACE and QIOS (under Grant Agreement n. 306576). The research leading to these results has received funding from the European Community’s Seventh Framework Programme (FP7/2007-2013) under grant agreement n211464-2 and it was further supported by the Villum Kann Rasmussen Centre of Excellence “NAMEC” under Contract No. 65286. We would like to thank Louise Jørgensen for cleanroom support and Martin Benjamin B. S. Larsen for support with the Raman measurements.

%\bibliographystyle{unsrt}
%\bibliography{/Users/sils/Documents/library}

\clearpage

%%%%%%%%%%%%%%%%%%%%%%%%%%%%%%%%%%%%%%%%%%%%%%%%%%%%%%%%%%%%%%%%%%%%%%%%%%%%%%%%%%%%%%%%%%%%%%%%

\setcounter{page}{0}%
\setcounter{figure}{0}%
\setcounter{equation}{0}%
\setcounter{section}{0}
\renewcommand \theequation {S\arabic{equation}}%
\renewcommand \thefigure {S\arabic{figure}}
\renewcommand \thesection {S \arabic{section}}
\renewcommand \thepage {S\arabic{page}}

\vspace{.2in}

%\bibliographystyle{plain}
%\renewcommand{\baselinestretch}{0.1}
%\begin{document}
\begin{widetext}

\thispagestyle{empty}

\begin{center}
\large{
\textbf{ Supplementary Information for \\ \small{Graphene on silicon nitride for optoelectromechanical micromembrane resonators}}}
\end{center}
\vspace{.2in}

%Intro		

\section{Electrostatic force constant}\label{sec:force_constant}
%\subsection{Symmetric coupling to electrodes}
In the case where the electrostatic coupling is symmetric with respect to the two electrode polarities, the force on a segment of membrane floating above the electrodes can be described by
\begin{equation}\label{eq:forceSI}
F = c A f(d) U^2
\end{equation}
with the area of the segment $A$, distance between fixed electrodes and membrane $d$, DC voltage $U$ between fixed electrodes, and an electrostatic force constant $c$; $f(d)$ is a scaling function with units of inverse area.

In writing the force law (\ref{eq:forceSI}) we have assumed the membrane dynamics to take place under fixed voltage conditions. This requires the voltage source to supply and absorb charge at the timescale on which the membrane motion modulates the capacitance. However, even if we operate in the opposite limit, i.e. fixed charge on the capacitor, the appropriate correction to the force law (\ref{eq:forceSI}) is negligible for our setup \cite{CouplingFrameworkInPrep}; this is due to stray capacitance (in the cables from the voltage source) that is much larger than that of the electrode-membrane arrangement. Intuitively, the large stray capacitance acts as a charge reservoir for the modulated electrode-membrane capacitor, effectively keeping the electrode voltage constant.

The equilibrium of forces for an infinitesimal piece of membrane with the area $dx \times dy$ and thickness $h$ is
\begin{equation}\label{eq:equationofmotion}
\sigma_0 h \nabla^2 w(x,y,t) - \rho h \frac{\partial^2 w}{\partial t^2}(x,y,t) + c U^2 f(d-w(x,y,t)) \xi(x,y) = 0
\end{equation}
with the displacement function $w(x,y,t)$, the tensile pre-stress $\sigma_0$, and the mass density $\rho$. $\xi(x,y)$ is a Heaviside step function (or a product of such functions) taking into account electrode gaps and edges, as well as the hole in the membrane metallization of some of the membranes used.
%\begin{equation}
%\xi(x,y) = H\left(\frac{x^2+y^2}{r^2}-1\right).
%\end{equation}
%The gaps between the electrodes are not taken into account in the model.

The deflection of a membrane can be described by
\begin{equation}\label{eq:drum_mode_expansion}
w(x,y,t) = \sum_{n=1}^\infty \sum_{m=1}^\infty A_{nm} \Phi_{n,m}(x,y) \, e^{i\omega t}
\end{equation}
with the mode shape function
\begin{equation}\label{eq:modeshape}
\Phi_{n,m}(x,y) = \sin\frac{n \pi x}{L_x} \sin\frac{m \pi y}{L_y}
\end{equation}

Considering the first order Taylor approximation of the electrostatic force (\ref{eq:forceSI}), the equation of motion can be written as
\begin{equation}
\sigma_0 h \nabla^2 w -  \rho h \frac{\partial^2 w}{\partial t^2} + \left(c U^2 f(d) - c U^2 f'(d) w \right)\xi(x,y) =0
\end{equation}
In a ''linear system'', the static force term $c U^2 f(d) \xi(x,y)$ causes a static deflection of the membrane. This static deflection does not influence the eigenfrequency and can thus be neglected. Following Galerkin's method, (\ref{eq:equationofmotion}) can be solved for the fundamental normal mode by multiplying it with $\Phi_{1,1}$ and integrating over the entire membrane area $A=L\times L$ with $L=L_x=L_y$
\begin{equation}
\iint\limits_A\!\left( \sigma_0 h \nabla^2 w + \rho h \omega^2 w - c U^2 f'(d) w\xi(x,y) \right) \Phi_{1,1} \dif x\,\dif y = 0
\end{equation}
which with (\ref{eq:drum_mode_expansion}) can be written as
\begin{equation}\label{eq:Galerkin}
-2\frac{\pi^2}{L^2}\sigma_0 h\!\iint\limits_A\!\Phi_{1,1}^2\dif x\,\dif y + \rho h \omega^2 \!\iint\limits_A\!\Phi_{1,1}^2\dif x\,\dif y - c U^2 f'(d) \!\iint\limits_A\!\Phi_{1,1}^2\xi(x,y)\dif x\,\dif y   = 0,
\end{equation}
where we have ignored the induced couplings to other membrane modes assumed to be weak.

The frequency can now be isolated from (\ref{eq:Galerkin})
\begin{equation}\label{eq:afterGalerkin}
\omega^2 = 2\pi^2 \frac{\sigma_0}{\rho}\frac{1}{L^2} + c\frac{U^2 f'(d)}{h \rho} \frac{\iint\limits_A\! \Phi_{1,1}^2\xi(x,y)\dif x\,\dif y}{\iint\limits_A\!\Phi_{1,1}^2\dif x\,\dif y}
\end{equation}
With the eigenfrequency $\omega_0$ of a membrane with zero voltage applied ($U=0$)
\begin{equation}
\omega_0 =\frac{\sqrt{2}\pi}{L}\sqrt{\frac{\sigma_0}{\rho}}
\end{equation}
and introducing the overlap factor $\eta_{1,1}$ between the membrane mode $\Phi_{1,1}$ and the electrode mask
\begin{equation}\label{eq:eta_sym}
\eta_{1,1} = \frac{\iint\limits_A\! \Phi_{1,1}^2\xi(x,y)\dif x\,\dif y}{\iint\limits_A\!\Phi_{1,1}^2\dif x\,\dif y},
\end{equation}
the first order Taylor approximation of $\omega$ becomes
\begin{equation}
\omega \approx \omega_0\left(1 + \frac{c}{2}\frac{U^2 f'(d)}{h \rho\omega_0^2} \eta_{1,1} \right)
\end{equation}
which results in a relative frequency shift of
\begin{equation}
\frac{\Delta\omega}{\omega_0} = \frac{c}{2}\frac{U^2 f'(d)}{h \rho\omega_0^2} \eta_{1,1}.
\end{equation}
For a membrane with a continuous electrode ($\xi =1 \Rightarrow \eta_{1,1}=1$) the frequency shift due to the electrostatic spring softening effect becomes
\begin{equation}
\frac{\Delta\omega}{\omega_0} = \frac{c}{2}\frac{U^2 f'(d)}{h \rho\omega_0^2}.
\end{equation}

In our experiments, the electrostatic force factor $c$ was determined by measuring the relative frequency shift of a membrane as a function of $U$. The measured data was fitted with
\begin{equation}\label{eq:softening}
\frac{\Delta\omega}{\omega_0} = - \alpha U^2
\end{equation}
and $c$ can then be obtained from
\begin{equation}\label{eq:c_exp_eta11}
c = 2 \alpha [-f'(d)]^{-1} h \rho \omega_0^2 \eta_{1,1}^{-1}
\end{equation}
or 
\begin{equation}
c = 2 \alpha [-f'(d)]^{-1} h \rho \omega_0^2
\end{equation}
for $\xi=1$.

\section{Theoretical force constant for a conductive membrane placed over coplanar quarter-segment electrodes}\label{sec:theory_c_4seg}
\subsection{Symmetrical coupling}
Dismissing fringe fields, the force between the plates of a parallel plate capacitor with distance $d$ is
\begin{equation}\label{eq:force_parallel}
F_{C}=\frac{1}{2}\varepsilon_0 \frac{A}{d^2}U^2
\end{equation}
where $\varepsilon_0$ is the vacuum permittivity. The potential of the floating conductive membrane is $U/2$, given that it is placed symmetrically in relation to the two electrode polarities. Thus, the force on a conductive membrane segment of area $A$ to coplanar quarter-segment electrodes is
\begin{equation}
F_{C}=\frac{1}{2}\varepsilon_0 \frac{A}{d^2}(U/2)^2
\end{equation}
and therewith the force constant (\ref{eq:forceSI}) becomes
\begin{equation}\label{eq:c_sym_theory}
c_{\text{sym}} = \frac{1}{8}\varepsilon_0,
\end{equation}
using the scaling function $f(d)=d^{-2}$.

\subsection{Hypothetical effects due to excess graphene}\label{sec:theory_c_4seg_asym}
In our experiments with SiN-G membranes on quarter-segment electrodes, it is possible that we may have additional capacitance contributions from the excess graphene outside the membrane area (as opposed to the case of SiN-Al membranes where the Al layer is strictly confined to within the SiN membrane area). Specifically, such extra contributions may occur in a highly asymmetric manner with regard to the two electrode polarities.

This prompts us to derive the theoretical value of $c$ for the case of highly asymmetric capacitance between the conductive membrane and the two electrode polarities. In this case the membrane will be essentially equipotential with one of the electrode polarities whereby the electrostatic force vanishes on membrane segments above electrodes of this polarity. Hence, under these circumstances, the membrane coupling to the electrodes is extremely asymmetric wrt.~the two polarities and $c$ acquires a dependence on position in the membrane plane: The force constant vanishes for the part of the membrane above electrodes with which it is equipotential, while for membrane segments above the other polarity, with which the potential difference is $U$, Eq.~(\ref{eq:force_parallel}) leads to
\begin{equation}\label{eq:c_asym_theory}
c_{\text{asym}} = \frac{1}{2}\varepsilon_0,
\end{equation}
which is four times the symmetrical value, Eq.~(\ref{eq:c_sym_theory}).

We will now discuss how the above effect, if present in our setup, would affect the force constants extracted from experiment. In the simple case where the membrane \emph{mode} is symmetric wrt.~the two electrode polarities (while the electrostatic force is not), the experimental force constant, as calculated using Eq.~(\ref{eq:c_exp_eta11}), samples the effective and ineffective electrodes equally, resulting in $c=\frac{1}{4}\varepsilon_0$ (given that the perfect conductor model is valid for SiN-G). This is a factor of 2 enhancement relative to the theoretical value for the case of symmetrical capacitances, Eq.~(\ref{eq:c_sym_theory}).

Due to a somewhat asymmetric electrode mask (featuring a centrally placed connection bar not shown in Fig. 1 of the main text) and, more importantly, the possibility of misalignment (estimated to be up to 25\% of the membrane side length), the simple result for the symmetrical membrane mode does not apply in general. In fact, by means of Eq.~(\ref{eq:eta_general_4seg}) presented below, we find that the addition of a large, asymmetrical capacitance wrt.~one polarity can enhance the fundamental mode frequency shift by a factor of $1.0-4.2$ for alignments within the cited range. By assuming a flat distribution within this range, we find that the effect would enhance the averaged force constant by a factor of 2.1.
A related mechanism, that would result in exactly the same enhancement as just described, is involuntary charging of the membrane by short-circuiting with one of the electrode polarities with $RC$ time much less than the membrane oscillation period $2\pi/\omega_{\text{mem}}$.

On the other hand, consider a similar short-circuiting but with $RC$ time longer than the membrane period while still much less than the scan time for the DC voltage $U$. In this situation the dynamics take place with fixed charge on the membrane and without access to the charge reservoir of the large stray capacitance as the latter now acts in series with the much smaller membrane-electrode capacitance (compare to the situation described in section \ref{sec:force_constant}). We find that the fundamental mode frequency shift can be modified by a factor of $0.8-2.5$ within the alignment range stated earlier. Assuming a flat alignment distribution once again, we find that the averaged force constant would be enhanced by a factor of 1.4 in case of such membrane charging.

Regarding SiN-G on quarter-segment electrodes, either of the above effects could serve to explain, at least partially, the large error bars of the average force constant (presented in Fig. 3c of the main text). 
Moreover, the effects of strongly asymmetrical capacitance and charging with short $RC$ time enhance the frequency shift in a way that is compatible with the average experimental force constant lying above the theoretical value for symmetrical coupling by a factor of 2.7.

\section{General asymmetric coupling to quarter-segment electrodes}\label{sec:Asym_coupling_4seg}
We discussed in section \ref{sec:theory_c_4seg} the two limits of coupling a conductive membrane to quarter-segment electrodes: Symmetric and strongly asymmetric. In calculating the experimental force constants presented in the main text for the quarter-segment geometry, we assume that the capacitance between the electrodes and the conductive layer on the membrane only has contributions from the membrane area (however, see remarks regarding excess graphene in section \ref{sec:theory_c_4seg_asym}).
To be able to account for electrode mask asymmetry and misalignment, we model the coupling as two capacitors in series: The capacitance between the positive electrode segments with the membrane in series with that between the negative electrodes and the membrane (note that $d$ is much smaller than the gap between fixed electrodes). This leads to the following generalization of the overlap factor $\eta$, Eq.~(\ref{eq:eta_sym}), for the quarter-segment geometry \cite{CouplingFrameworkInPrep}:
\begin{equation}\label{eq:eta_general_4seg}
\eta_{n,m} = 4 \left[     \frac{O_{+}^{(2)}}{\left(1+\frac{O_{+}^{(0)}}{O_{-}^{(0)}}\right)^2} \\
				+\frac{O_{-}^{(2)}}{\left(1+\frac{O_{-}^{(0)}}{O_{+}^{(0)}}\right)^2} \\
				- \frac{1}{O_{+}^{(0)} + O_{-}^{(0)}} \\
				\left(\frac{\frac{O_{+}^{(1)}}{O_{+}^{(0)}} - \frac{O_{-}^{(1)}}{O_{-}^{(0)}}}{\frac{1}{O_{+}^{(0)}} + \frac{1}{O_{-}^{(0)}}}  \right)^2
			   \right],
\end{equation}
where we have introduced the symbol
\begin{equation}
O_{i}^{(j)} \equiv \frac{\iint\limits_{A_{i}}\! \Phi_{n,m}^j\xi(x,y)\dif x\,\dif y}{\left(\iint\limits_A\!\Phi_{n,m}^2\dif x\,\dif y\right)^{j/2}},
\end{equation}
with $i\in\{+,-\}$ and $A_i$ being the area of the membrane above electrodes of polarity $i$. The terms in Eq.~(\ref{eq:eta_general_4seg}) can be interpreted as follows: The first (second) term represents the frequency shift contribution from the membrane area above positive (negative) electrodes with the local potential difference fixed at its equilibrium value, whereas the third term is a correction due to the modulations of the membrane potential around the equilibrium value. We use Eq.~(\ref{eq:eta_general_4seg}) in calculating experimental $c$ values in order to account for gaps, edges and a slight asymmetry in the electrode mask as well as the central hole in the Al layer of the SiN-Al membranes. The formula was also used to estimate the uncertainty in $c$ due to lateral misalignment of the membrane wrt.~the electrodes (estimated to be up to 25\% of the membrane side length).

\section{Raman characterization of SiN-G membranes}
The graphene on the SiN-G membranes was characterized by Raman spectroscopy. The graphene flakes are covering the entire SiN membranes plus a large part of the surrounding frame. In Fig.~\ref{fig:raman} the Raman spectra of a SiN-G membrane chip is shown. The graphene spectra clearly show the typical spectrum of single layer graphene with a 2D-peak having double the amplitude of the G-peak. In all 22 point measurements we made the D-peak at 1300-1400~cm$^-1$ was missing which indicates that the graphene layers is of good quality with probably only few defect. 
\begin{figure}
\centering
\includegraphics[width=0.7\textwidth]{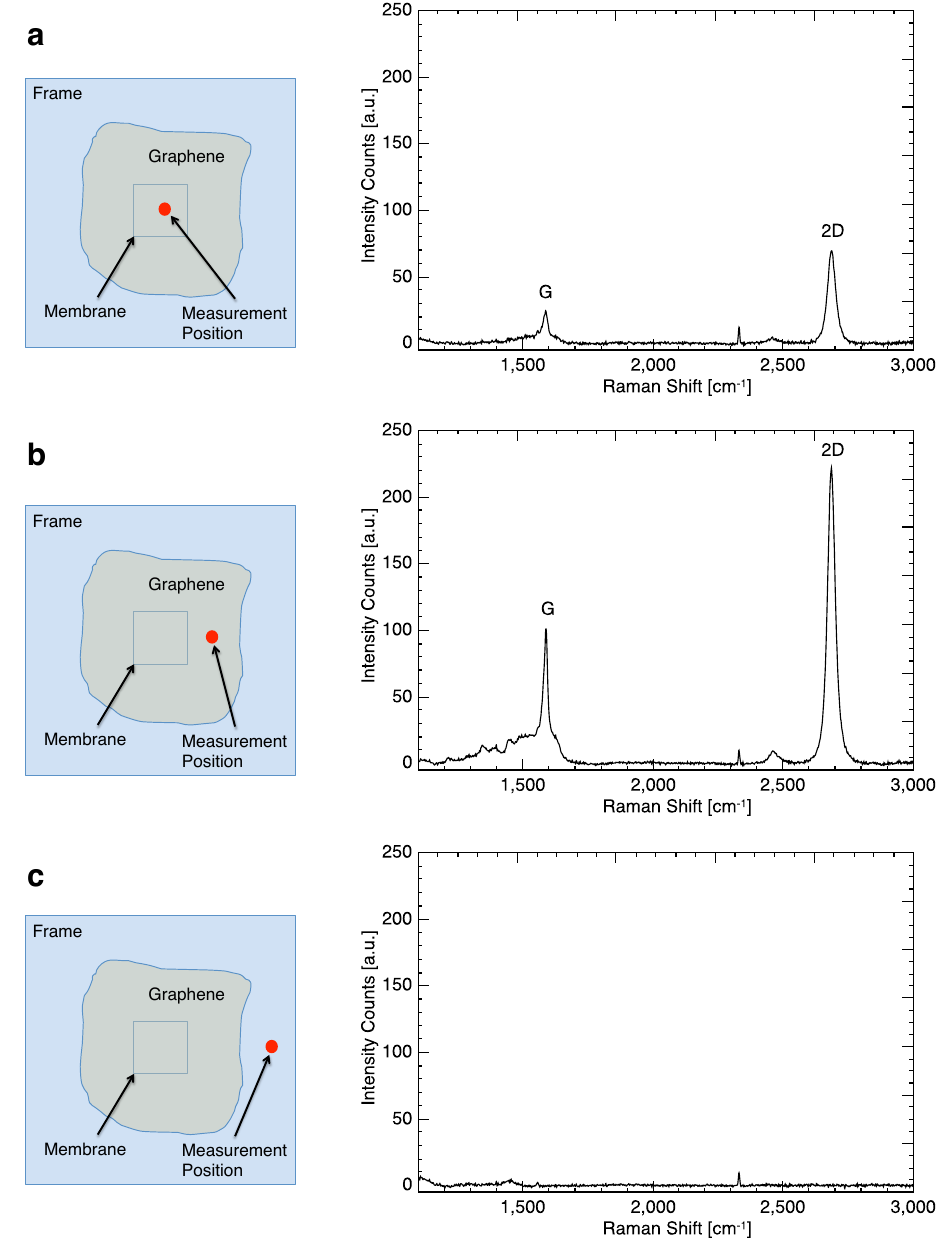}
\caption{Raman spectra of a SiN-G chip measured on a) the graphene covering the SiN membrane, b) the graphene covering the frame, and c) the blank frame. The spectra were measured with a 532~nm laser with 10~mW power and 15 exposures with an exposure time of 15~s.}
\label{fig:raman}
\end{figure}

\section{Charging of bare SiN membranes}
In the case of charging, a linear force adds up to the quadratic electrostatic force. Thus, a linear term has to be added to the relative frequency shift as a function of DC voltage (\ref{eq:softening})
\begin{equation}\label{eq:softeningwithcharge}
\frac{\Delta \omega}{\omega} = -\alpha U^2_{DC} - \beta U_{DC}.
\end{equation}
Fig.~\ref{fig:chargingSI}a shows the charging induced hysteresis recorded during the charging experiment presented in Fig.~4a in the main document. In the experiment, the DC voltage was swept 4 times between 0 and +30~V. From sweep to sweep the spring softening becomes continuously stronger. This is a clear sign for the accumulation of charge on the bare SiN membrane. In the 5th sweep, the voltage is swept from 0 to -30~V and a spring hardening is observed. In Fig.~\ref{fig:charging}b, the initial sweep is fitted with the charge free model (\ref{eq:softening}) and the last continuous sweeps 4+5 are fitted with (\ref{eq:softeningwithcharge}) which contributes for a linear force from free charge on the membrane. The fit is of good quality and strongly supports the hypothesis of charge accumulation on the membrane. The force constant of the linear force due to free charge is one order of magnitude larger than the force constant of the electrostatic force. Hence, free charge on the membrane significantly contributes to the electromechanical membrane behavior especially at low DC voltages.
\begin{figure}
\centering
\includegraphics[width=0.7\textwidth]{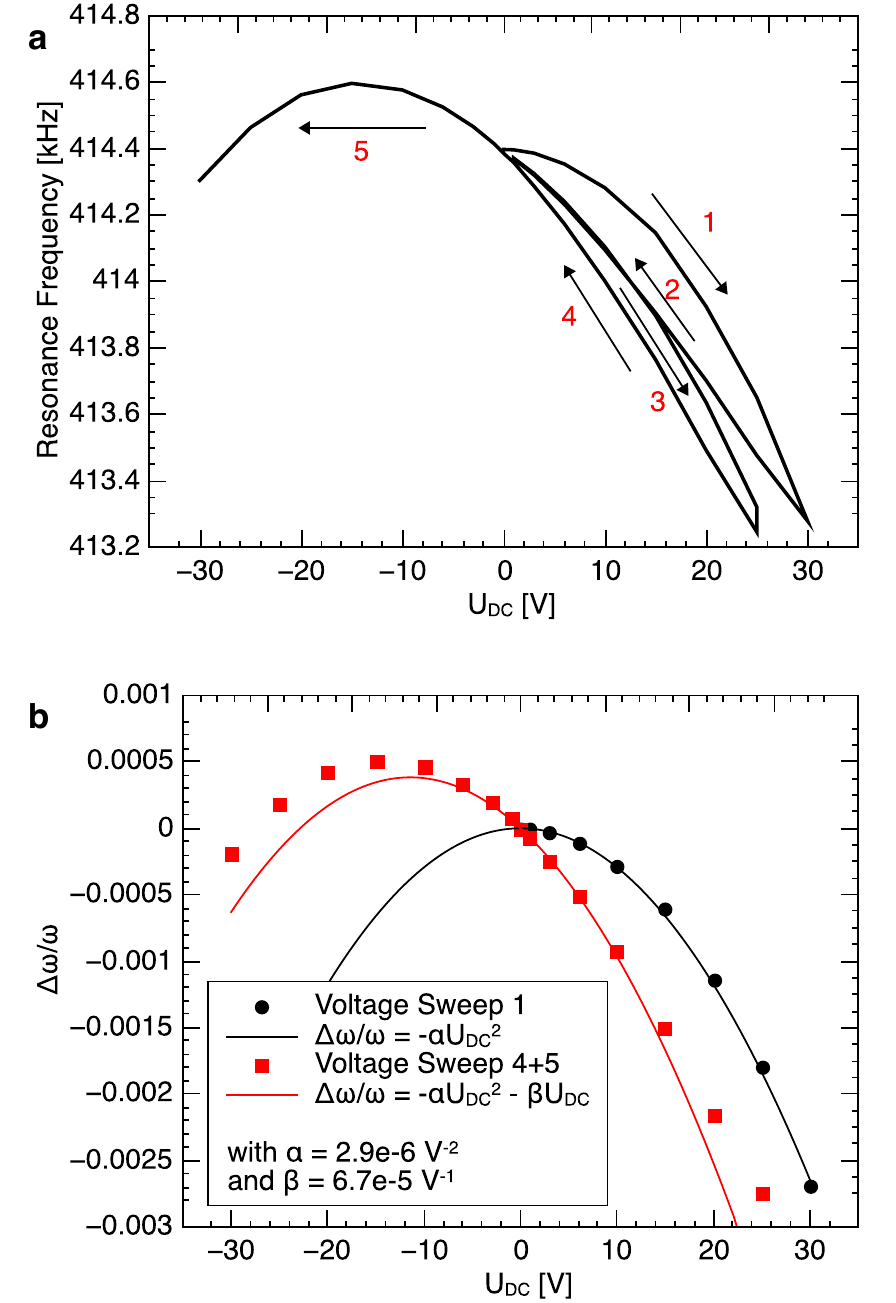}
\caption{a) Detail of the charging experiment shown in Fig.~4a in the main document. The graph shows the response of a bare SiN membrane (1$\times$1~mm$^2$, d = 4.0~$\mu$m) on interdigitated electrodes; electrode fingers are 4~$\mu$m wide with a gap of 2~$\mu$m between the fingers. The voltage was swept 4 times between 0 to +30~V and backwards. In the end the DC voltage was swept from 0 to -30~V. b) Fig.~\ref{fig:charging}b shows the fitting of the initial voltage sweep 1 with (\ref{eq:softening}) and the last sweep 4+5 with (\ref{eq:softeningwithcharge}). The fitting parameters are for sweep 1: $\alpha = 2.9\times10{-6}$~V$^2$, and for sweep 4+5: $\beta = 6.7\times10{-5}$~V$^{-1}$ while $\alpha$ was kept constant at $\alpha = 2.9\times10{-6}$~V$^2$.}
\label{fig:chargingSI}
\end{figure}

%\bibliographystyle{unsrt}
%\bibliography{EZ}

\end{widetext}
\end{document}